\documentclass[aps,prl,floats,fleqn]{revtex}
\usepackage[latin1]{inputenc}
\usepackage{babel}
\usepackage{graphics}

\makeatletter

\newcommand{\LyX}{L\kern-.1667em\lower.25em\hbox{Y}\kern-.125emX\spacefactor1000}

\newcommand{\boldsymbol}[1]
{\mbox{\boldmath $#1$}}

\usepackage{epsf}
\usepackage{rotate}

\def \point{\;\;.}
\makeatother

\begin{document}

\wideabs{

\title{SO(N) symmetries in the two--chain model of correlated fermions}

\author{H.J. Schulz \\
Laboratoire de Physique des Solides, Université Paris--Sud \\
91405 Orsay,France }

\maketitle
\begin{abstract}
The two--chain model of correlated fermions, appropriate e.g. for doped spin
ladders or carbon nanotubes, is investigated in the weak--interaction limit.
It is shown that the model scales to situations with high internal symmetries,
generically SO(6), whereas on critical lines there is SO(5) or SO(3) symmetry.
Consequences for the level spectroscopy of the model are discussed. In particular,
the ``\( \pi  \)--mode'' is shown to be sixfold degenerate and contains both
spin and charge excitations.
\end{abstract}
}\draft 

Models of two coupled chains of interacting fermions are of interest for a variety
of reasons: they are directly applicable to interesting experimental systems
like doped spin ladders\cite{uehara96} or carbon nanotubes\cite{kane96}, and
they also provide a first step for the understanding of the transition from
the well--understood single chain case to the area of two--dimensional correlated
fermions.\cite{dagotto_ladders} In particular, both analytical \cite{fabrizio_2chain,schulz_2chain,balents_2chain}
and numerical \cite{dagotto_tjladder,hayward_2chain,noack_2chain} work show
that the two--chain system with purely repulsive interactions is a d--type superconductor
whereas the single chain is antiferromagnetic. This provides a potential explanation
for the observation of superconductivity in a doped spin ladder,\cite{uehara96}
and is probably the best--established case of correlation--induced superconductivity.

In recent work\cite{lin98} it has been shown that the Mott--insulating half--filled
two--chain model has a very big SO(8) symmetry in the weak--correlation limit,
leading to a number of interesting predictions. Away from half--filling the
SO(8) symmetry is broken and the system becomes conducting. Here I will study
the symmetries of the conducting state away from half--filling and show that
generically a still large SO(6) (=SU(4)) symmetry remains. On critical lines
the symmetry is lower, SO(3) or SO(5). There are exact solutions corresponding
to these symmetric models, \cite{zamolodchikov79,andrei80} and from those it
is possible to obtain quite detailed information about the excitation spectrum
and in particular the appearance and location of bound states. The present work
also provides a microscopic extension of the SO(5) model of correlated fermions
\cite{zhang97} and the associated ladder models.\cite{scalapino98}

I consider the standard two--chain model of interacting fermions.\cite{fabrizio_2chain,schulz_2chain,balents_2chain}
After linearizing the electron dispersion around the Fermi energy, the kinetic
energy density takes the form ,
\begin{equation}
\label{eq:h0}
h_{0}=-iv_{F}\sum _{s,q=0,\pi }\{\psi ^{\dagger }_{Rsq}\partial _{x}\psi ^{\phantom {\dagger }}_{Rsq}-\psi ^{\dagger }_{Lsq}\partial _{x}\psi ^{\phantom {\dagger }}_{Lsq}\}
\end{equation}

\noindent Here \( q=0 \) (\( \pi  \)) for the (anti--)bonding bands split
by the interchain hopping integral \( t_{\perp } \), \( s \) is the spin,
\( R \) (\( L \)) refers to the right-- (left--)going fermions, and \( \psi _{R(L)sq} \)
are the corresponding fermion field operators.  

Important interaction effects arise from processes where both the two incoming
and the two outgoing particles are all at the Fermi energy. The Hamiltonian
density of the interaction then is  

\noindent 
\begin{eqnarray}
h_{int} & = & g_{1,abcd}\psi ^{\dagger }_{Rsa}\psi ^{\dagger }_{Ltb}\psi ^{\phantom {\dagger }}_{Rtc}\psi _{Lsd}^{\phantom {\dagger }}\nonumber \\
 & + & g_{2,abcd}\psi ^{\dagger }_{Rsa}\psi ^{\dagger }_{Ltb}\psi ^{\phantom {\dagger }}_{Ltc}\psi _{Rsd}^{\phantom {\dagger }}\label{eq:h1} 
\end{eqnarray}
These terms describe backward (\( g_{1} \)) and forward (\( g_{2} \)) scattering
between particles at the Fermi surface. Summation over the spin indices \( s,t \)
and the band indices \( a,b,c,d \) is understood. The interactions allowed
by momentum and energy conservation are \( g_{1,0000}=g_{1,\pi \pi \pi \pi }\equiv g_{11} \)
, \( g_{1,00\pi \pi }=g_{1,\pi \pi 00}\equiv g_{12} \), \( g_{1,0\pi 0\pi }=g_{1,\pi 0\pi 0}\equiv g_{13} \),
\( g_{2,0000}=g_{2,\pi \pi \pi \pi }\equiv g_{21} \), \( g_{2,0\pi \pi 0}=g_{2,\pi 00\pi }\equiv g_{22} \),
\( g_{2,00\pi \pi }=g_{2,\pi \pi 00}\equiv g_{23} \).

The one--loop renormalization group equations for the coupling constants under
a change of energy cutoff \( \Lambda \rightarrow \Lambda e^{-l} \) have been
found previously \cite{varma_2chain} and are given by: 
\begin{eqnarray}
g_{11}' & = & -2g_{11}^{2}-2g_{12}g_{23}\nonumber \label{t} \\
g_{12}' & = & -2(g_{12}g_{21}+g_{11}g_{23}+2g_{12}g_{13}-g_{13}g_{23}-g_{12}g_{22})\nonumber \label{eq:2:rg2} \\
g_{13}' & = & -2g_{12}^{2}+2g_{12}g_{23}-2g_{13}^{2}\nonumber \label{eq:2:rg3} \\
g_{21}' & = & -g_{11}^{2}-g_{12}^{2}-g_{23}^{2}\nonumber \label{eq:2:rg4} \\
g_{22}' & = & -g_{13}^{2}+g_{23}^{2}\nonumber \label{eq:2:rg5} \\
g_{23}' & = & -2g_{11}g_{12}-2(g_{21}-g_{22})g_{23}\label{eq:2:rg6} 
\end{eqnarray}

\noindent Here \( g_{ij}'\equiv 2\pi v_{F}\frac{dg_{ij}}{dl} \)

Abelian bosonization is used to write the model in terms of symmetric and antisymmetric
charge (\( \nu =\rho \pm  \)) and spin (\( \nu =\sigma \pm  \)) fields \( \phi _{\nu } \)
and their conjugate momentum densities \( \Pi _{\nu } \). The total Hamiltonian
becomes the sum of a term bilinear in boson fields (\( H_{b} \)) and a term
containing nonlinear interactions \( H_{\mathrm{nl}} \), with \( H_{b}=H_{\rho +}+H_{\rho -}+H_{\sigma +}+H_{\sigma -} \),

\[
H_{\nu }=\frac{\pi v_{F}}{2}\int dx\left\{ \left( 1+M_{\nu }\right) \Pi _{\nu }^{2}+\frac{1}{\pi ^{2}}\left( 1-M_{\nu }\right) (\partial _{x}\phi _{\nu })^{2}\right\} ,\]
and \( M_{\sigma \pm }=(g_{11}\pm g_{13})/2\pi v_{F} \) ,\( M_{\rho \pm }=M_{\sigma \pm }-(g_{21}\pm g_{22})/\pi v_{F} \).
The interaction term is
\begin{eqnarray}
H_{\mathrm{nl}} & = & \frac{1}{(\pi \alpha )^{2}}\int dx\{(g_{23}-g_{12})\cos 2\theta _{\rho -}\cos 2\theta _{\sigma -}\nonumber \\
 & + & g_{23}\cos 2\theta _{\rho -}\cos 2\phi _{\sigma -}+g_{11}\cos 2\phi _{\sigma +}\cos 2\phi _{\sigma -}\nonumber \\
 & + & g_{12}\cos 2\theta _{\rho -}\cos 2\phi _{\sigma +}-g_{13}\cos 2\phi _{\sigma +}\cos 2\theta _{\sigma -}\},\label{eq:hnl} 
\end{eqnarray}

\noindent where \( \theta _{\nu } \) is the field dual to \( \phi _{\nu } \):
\( \Pi _{\nu }=\partial _{x}\theta _{\nu }/\pi  \).

\begin{figure}
{\centering \resizebox*{0.35\textwidth}{!}{\includegraphics{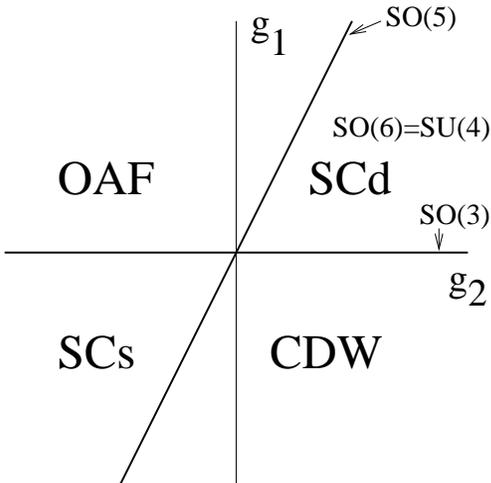}} \par}

\caption{``Phase diagram'' in the \protect\( g_{1}-g_{2}\protect \) plane (\protect\( g_{1i}\equiv g_{1},g_{2i}\equiv g_{2}\protect \)),
indicating the dominant fluctuation. CDW: charge--density wave, SCd: d--wave
superconductor, OAF: orbital antiferromagnet, SCs: s--wave superconductor. On
the critical lines \protect\( g_{2}=0\protect \) and \protect\( g_{1}=2g_{2}\protect \)
additional critical fluctuations occur.\label{f:2.2}}
\end{figure} It is clearly interesting to understand the energetics and quantum numbers
of the low--lying excited states of the model, in particular in order to understand
the relation to numerical or experimental results. I start by investigating
the generic regions of the phase diagram, fig.\ref{f:2.2}, i.e. away from the
critical lines \( g_{1}=0 \), \( g_{1}-2g_{2} \). In particular, in the \( SCd \)
region, important for Hubbard--like bare interactions, a numerical investigation
shows that the asymptotic behavior near the divergence (\( l\rightarrow l_{c} \))
is given by \( \tilde{g}_{ij}(l)=\widetilde{g}_{ij}^{0}/(l_{c}-l) \), with
the values 
\begin{eqnarray}
-\tilde{g}^{0}_{11}=\tilde{g}^{0}_{12}=\tilde{g}^{0}_{23}=1/4,\quad \tilde{g}^{0}_{13}=0, &  & \nonumber \\
\quad \tilde{g}^{0}_{21}=-3/16,\quad \tilde{g}^{0}_{22}=1/16\point  &  & \label{eq:asym2} 
\end{eqnarray}

\noindent One can note that eq.(\ref{eq:asym2}) is an exact solution of the
renormalization group equations (\ref{eq:2:rg6}) for a special set of initial
conditions. For more general interactions, in particular the Hubbard model,
the solution of the renormalization group equations tends asymptotically for
\( l\rightarrow l_{c} \) towards this solution, i.e. eq.(\ref{eq:asym2}) is
the ``fixed ray'' towards which all solutions of the renormalization group
equations scale in the \( SCd \) part of the phase diagram.  

Interestingly, for the parameter values eq.(\ref{eq:asym2}) the model exhibits
a high but somewhat hidden symmetry: after the particle--hole transformation
on the left--going branch \( \psi ^{\dagger }_{Lsq}\leftrightarrow \psi _{L-sq} \)
takes the \( U(4) \) invariant form 
\begin{equation}
\label{eq:u2a}
H=H_{0}-g\int dx(\frac{1}{4}\rho _{R}\rho _{L}+\psi ^{\dagger }_{Rsq}\psi ^{\phantom {\dagger }}_{Lsq}\psi ^{\dagger }_{Ltp}\psi ^{\phantom {\dagger }}_{Rtp}),
\end{equation}

\noindent where \( \rho _{\alpha }=\psi ^{\dagger }_{\alpha sq}\psi ^{\phantom {\dagger }}_{\alpha sq} \).
This is the \emph{chiral} Gross--Neveu model, for which an exact Bethe ansatz
solution exists.\cite{andrei80} 

Another transformation, which will be more generally useful is obtained by refermionizing
the bosonized Hamiltonian, eq.(\ref{eq:hnl}), by introducing three fermion
fields \( \psi _{1,2,3} \) corresponding to the \( (\rho -) \), \( (\sigma +) \),
and \( (\sigma -) \) boson fields. After a duality transformation \( \phi _{\rho -}\leftrightarrow \theta _{\rho -} \)
and writing the Dirac fermions in terms of hermitian (Majorana) fields \( \chi  \)
as \( \psi _{\alpha i}=\chi _{\alpha ,2i-1}+i\chi _{\alpha ,2i} \) the Hamiltonian
in this subspace then takes the form 

\noindent 
\begin{equation}
\label{eq:so6}
H=H_{0}+2g\int dx\, \left( \sum _{i=1}^{N=6}\chi _{Ri}\chi _{Li}\right) ^{2},
\end{equation}

\noindent This is the well--known Hamiltonian of the Gross--Neveu model\cite{gross74}
which has a SO(6) symmetry. Due to the equality of the Lie algebras so(6) and
su(4), the equivalence of the chiral SU(4) and the (non--chiral) SO(6) models
may seem quite natural. One should however notice that the nonchiral model has
a double--degenerate symmetry--broken ground state,\cite{gross74,shankar78}
whereas there is no symmetry breaking in the chiral model. This will lead to
interesting consequences for the quantum numbers of the excited states. It should
be emphasized that eq.(\ref{eq:so6}) describes the massive part of the spectrum
and that the total charge mode \( \rho + \) remains massless in all cases and
therefore contributes to correlation functions in an important way.

We can now use known results for the excitation spectrum of the SO(N) models\cite{zamolodchikov79,andrei80}
to obtain results for the excitations of the two--chain model. For the generic
SU(4) case the elementary excitations are single fermions and holes with mass
gap \( m_{0}\propto e^{-1/g} \). Due to the particle--hole transformation leading
to eq.(\ref{eq:u2a}), right--going fermions transform as the fundamental \( \textbf {4} \)
representation, whereas left--going fermions transform as the conjugate \( \boldsymbol {\bar{4}} \).
In the SO(6) language these fermionic excitations are kinks separating the two
degenerate ground states,\cite{shankar78} whereas the SO(6) fermions correspond
to two--particle excitations in the original model.

From the single--fermion gap \( m_{0} \) one expects a gap \( 2m_{0} \) in
two--particle excitations. This is in fact correct for all excitations created
by operators of the form \( \psi ^{\dagger }_{Rsq}\psi ^{\dagger }_{Ltq'} \).
Consequently, all pairing correlation functions, irrespective of their spin
(singlet or triplet) or orbital structure, decay exponentially, with \emph{the
same correlation length} \( \xi =v_{F}/2m_{0} \). The one important exception
of course is the ``d--wave'' singlet channel, which has purely algebraic decay.
In the group theoretically language this corresponds to the decomposition \( \textbf {4}\otimes \boldsymbol {\bar{4}}={\textbf {1}}\oplus {\textbf {15}} \),
the \( \textbf {1} \) being the d--wave superconductor.  

More structure arises in the particle--hole channel, created by operators of
the type \( \psi ^{\dagger }_{Rsq}\psi ^{\phantom {\dagger }}_{Ltq'} \). Here
the relevant decomposition is \( {\textbf {4}}\otimes {\textbf {4}}={\textbf {6}}\oplus {\textbf {10}} \),
the \( \textbf {6} \) being the antisymmetric rank--2 tensor. Remarkably, the
exact solution shows that particle--hole excitations in this channel are bound
and appear at energy \( \sqrt{2}m_{0} \), rather than \( 2m_{0} \). The operators
creating these bound states are the spin triplet \( \psi ^{\dagger }_{Rs0}(\sigma _{\alpha })_{st}\psi ^{\phantom {\dagger }}_{Lt\pi }+(0\leftrightarrow \pi ) \)
which represents an \( 2k_{F} \) spin density wave which is out of phase between
the two chains, and the ``orbital triplet'' \( \psi ^{\dagger }_{Rsq}(\sigma _{\alpha })_{qq'}\psi ^{\phantom {\dagger }}_{Lsq'} \)
which contains a charge density wave which is in phase between the two chains
and the orbital antiferromagnet (OAF, or staggered flux phase). From the \( SU(4) \)
symmetry we can conclude that the gaps in these six channels are identical.
It is in particular remarkable that the lowest spin and charge gap appear, due
to bound state formation, at only \( \sqrt{2} \) times the single--particle
gap. One should notice that contrary to the triply--degenerate antiferromagnetic
\( \pi  \)--mode of the SO(5) model of correlated fermions \cite{zhang97},
we here have an \emph{extended \( \pi  \)--mode with sixfold degeneracy} and
which also contains charge excitations. The remaining particle--hole excitations
(the \( \textbf {10} \)) all have excitation energy \( 2m_{0} \).  

I now turn to the critical lines in fig.\ref{f:2.2}. Both are of Ising type.
For \( g_{1}=0 \) the asymptotics of the flow is now given by \( \tilde{g}^{0}_{1i}=0,\tilde{g}^{0}_{21}=-\tilde{g}^{0}_{22}=\tilde{g}^{0}_{23}/2=1/4. \)
One can now follow the same logic as before. However, here the self--dual combination
\( \cos 2\phi _{\sigma -}+\cos 2\theta _{\sigma -}\rightarrow \chi _{R3}\chi _{L3} \)
appears, and consequently \( \chi _{\alpha 4} \) decouples and remains massless,
together with the two boson fields \( \phi _{\nu +} \). The fermionic model
for the massive degrees of freedom then is given by eq.(\ref{eq:so6}), however
now with \( N=3 \). In this case the model has an \( \textrm{SU}(2)_{R}\times \textrm{ SU}(2)_{L}\textrm{ }\times \textrm{ SU}(2) \)
symmetry, where the first two subgroups represent spin rotations for the right-
and left--moving electrons (these are independent because \( g_{1j}=0 \)),
whereas the last \( \textrm{SU}(2) \) acts in an ``orbital'' space spanned
by the spinors \( \Psi _{Rs}=(\psi _{Rs0},\psi _{Rs\pi })^{T} \) and \( \Psi _{Ls}=(\psi _{Ls\pi },\psi _{Ls0})^{T} \).

There is a gap \( m_{0} \) in the single--particle (kink) excitation spectrum.
The two--particle and particle--hole spectra can be analyzed rewriting the \( \rho - \)
and \( \sigma - \) fields in terms of Ising model variables.\cite{shelton96}
This then shows that excitations of orbital singlet type are massless. In the
particle--particle channel this leads to powerlaw decay of d--wave pairing correlations,
both in the spin singlet and triplet channels. In the particle--hole channel,
the powerlaw decay occurs for CDW\( _{\pi } \) and SDW\( _{\pi } \) correlations
which are out of phase between the two chains. The exponent of the powerlaw
is entirely determined by the total charge degrees of freedom (\( \rho + \)).
For \( N\le 4 \) the Gross--Neveu model does not show two--kink bound states.\cite{zamolodchikov79}
Consequently, one easily checks that all other particle--hole or two--particle
excitations have the same gap \( 2m_{0} \), and the corresponding correlation
functions decay exponentially with distance. These states are all triplets under
the orbital SU(2).  

Finally, on the second critical line \( g_{1}=2g_{2} \) there is again a self--duality
in the \( \sigma - \) sector, but now the \( \sigma + \) degrees of freedom
also have a gap. Here \( \tilde{g}_{11}^{0}=\tilde{g}_{13}^{0}=\tilde{g}_{21}^{0}=-\tilde{g}_{23}^{0}=-1/6,\tilde{g}_{12}^{0}=1/3,\tilde{g}_{22}^{0}=0. \)
We again find a Gross--Neveu model, but now with \( N=5 \). The kink excitations
now form a quartet, leading again to a gap \( m_{0} \) in the single particle
Green functions. Arranging these quartets as \( \Psi _{R}=(\psi _{R\uparrow 0},\psi _{R\downarrow 0},i\psi _{R\uparrow \pi },i\psi _{R\downarrow \pi })^{T} \),
and \( \Psi _{L}=(i\psi _{L\uparrow \pi },i\psi _{L\downarrow \pi },\psi _{L\uparrow 0},\psi _{L\downarrow 0})^{T} \),
these transform as SO(5) spinors, and the Hamiltonian with these coupling constants
given by then is explicitly SO(5) invariant.  

Contrary to the SO(3) case, there are now two--kink bound states at energy \( \sqrt{3}m_{0} \),
in addition to unbound two--kink states at \( 2m_{0} \). These bound states
occur in the vector (\textbf{5}) part of the SO(5) decomposition \( {\textbf {4}}\otimes {\textbf {4}}={\textbf {1}}\oplus {\textbf {5}}\oplus {\textbf {10}} \).
In particular, in the particle--hole channel, the scalar (\textbf{1}) excitations
are massless and occur in the OAF channel, and the bound vector excitations
are the same as those in the particle--hole channel in the SU(4) case, with
OAF removed. All other particle--hole excitations, occurring in the antisymmetric
tensor (\textbf{10}) representation, have mass \( 2m_{0} \). In the particle--particle
channel the same decomposition applies, however now the massless scalar corresponds
to d--wave singlet pairing, the bound vector consists of s--wave triplet pairing
and two singlet paired states with transverse momentum \( \pi  \), and the
\textbf{10} contains all other pairing states.\begin{figure*}
{\centering \resizebox*{0.65\textwidth}{!}{\includegraphics{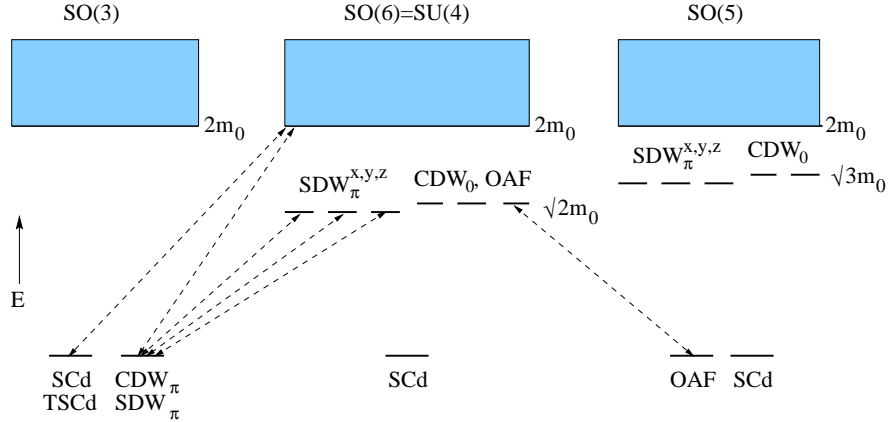}} \par}

\caption{Energy levels of the SO(N) models for different particle--hole and particle--particle
states in the region \protect\( 0\le g_{1}\le 2g_{2}\protect \) of the phase
diagram, fig.\ref{f:2.2}. Horizontal lines indicate discrete energy levels,
shaded areas two--particle continua. In the SO(5) case, the bound state at \protect\( \sqrt{3}m_{0}\protect \)
also appears in the two--particle channel, as discussed in the text. Labels
indicate in which correlation function the corresponding level appears. The
lowest levels in each column indicate massless correlations. TSCd: triplet d--wave
superconductor; (S)CDW\protect\( _{q}\protect \): (spin) charge density wave
in phase (\protect\( q=0\protect \)) and out of phase (\protect\( q=\pi \protect \))
between chains.\label{f:spectrum}}
\end{figure*}   

The energy levels in the interesting region \( 0<g_{1}<2g_{2} \), corresponding
to repulsive interactions, and on the adjacent critical lines are shown in fig.\ref{f:spectrum}.
In the SO(N) language, these are the discrete energy levels at \( q=0 \) or
\( q=2k_{F} \), however in our original model the presence of the massless
\( \rho + \) mode transforms these into lower edges of continua. Near this
edge, the correlation functions have singularities of the form \( (\omega -\omega _{0})^{-\nu } \),
where in weak coupling \( \nu \approx 1/2 \) and for the lowest (massless)
levels \( \omega _{0}=0 \) and \( \nu \approx 3/2 \). 

The full symmetries are only exact for asymptotically weak coupling. At any
finite coupling, corrections to the asymptotics of eqs.(\ref{eq:2:rg6}) will
lead to splittings in some multiplets. This is schematically indicated for the
bound multiplets in fig.\ref{f:spectrum}. A more dramatic consequence of the
corrections occurs in the close vicinity of the critical lines: for example,
in order to obtain the massless CDW\( _{\pi } \) correlations of the SO(3)
case, the correction terms have to form a bound state out of the continuum of
the SO(6) spectrum and to ``pull it down'' all the way to masslessness as
the SO(3) critical line is approached (as indicated by the broken lines in fig.\ref{f:spectrum}).
This is in particular of interest for the case of very weakly coupled chains
where renormalizations from the one--dimensional region \( E>t_{\perp } \)
lead to initial conditions for eqs.(\ref{eq:2:rg6}) obeying \( g_{1i}\ll g_{2i} \).
A detailed discussion is beyond the scope of the present paper.

Other parts of the phase diagram, fig.\ref{f:2.2}, can be discussed along similar
lines. Away from the critical lines one always finds an SO(6) symmetry. Of some
interest is the SCs region, which contains the Hubbard model with attractive
interaction which is numerically quite accessible and might serve as a testing
ground. Here again a sixfold degenerate low--lying state exists, however now
composed of the three \( CDW_{0,\pi } \) states and the spin nematic (the spin
triplet analogue of the orbital antiferromagnet).

In conclusion, I have shown that in the weak--interaction limit the generic
doped spin ladder model exhibits an SO(6) symmetry in its spectrum, weakly broken
by corrections to scaling. The most direct consequence is that there is a sixfold
degenerate \emph{extended \( \pi  \)--mode} in the d--wave superconducting
state which contains both spin and charge excitations. This is to be compared
with the SO(5) model \cite{zhang97} (where the symmetry is broken to SO(3)
by doping) and only a triply degenerate spin \( \pi  \)--mode is predicted.
Approximately equal charge and spin excitation energies have been observed in
numerical calculations \cite{poilblanc95} on t--J ladders, providing an indication
that the present results may be relevant even for strongly correlated systems.
To which extent the present results can be generalized to genuinely two--dimensional
situations remains to be studied.

\bibliographystyle{prsty}
\bibliography{revues,1dtheory,1dexp,2dtheory}

\end{document}